# Against Absolute Actualization: Three "Non-Localities" and Failure of Model-External Randomness made easy with Many-Worlds Models including Stronger Bell-Violation and Correct QM Probability

Sascha Vongehr[a]

College of Modern Engineering and Applied Sciences, Nanjing University, & Department of Philosophy, Nanjing University, Jiangsu 210093, P.R. China

Experimental violation of Bell-inequalities proves actualization of many futures (~ many-worlds); I show that this is not mere interpretation. To show this self-contained pedagogically (~ didactically), I resolve the Einstein-Podolsky-Rosen (EPR) paradox by starting with a visually intuitive *non*-quantum many-worlds model that already has 'apparent non-locality'. A modification leads to a classical-to-quantum transition model. 'Model-*ex*ternal randomness' (a ghost outside the universe throwing a pebble on state-space) stays unchanged, but the modeled observers witness strong Bell-violation. I derive the quantum probability $P$ from classical-to-quantum consistency. Model-internal probability (~ subjective Bayesianism) is derived as a measure of surprise (~ Deutsch's rational expectation). The model shows how absolute actualization, say by real hidden variables, fails. Models with $P$ and standard Bell-violation are supplied for completeness.

The transition model is the first touchable, interactive science-outreach exhibit teaching correct quantum physics. The discussion reinterprets the transition model, defends Einstein-locality, rejects "realism" etc., to the conclusion that the models teach the *necessity* of "many worlds/minds", rejecting probability concepts with 'random randomness' regress. The models teach how EPR empirically excludes certain realisms.

**Highlights:**
- Visually intuitive resolution of EPR paradox preserving Einstein-locality
- Clarifies three usually confused concepts of "non-locality"
- Clarifies that many-worlds models can be non-quantum
- Replaces subjective/objective by model-internal/external chance applicable to nested models
- Relative actualization (failure of absolute actualization), visually via pointers in the model
- Self-contained derivation of quantum frequencies from classical-to-quantum consistency
- Announcement of the first ever touchable, functioning exhibit of a quantum model
- In depth "philosophical" justification in separate discussion part (Supplemental Material)

---

[a] Corresponding author's electronic mail: vongehr@usc.edu

# 1 Introduction

Quantum physics refutes certain realisms empirically by proving "quantum non-locality", an often misinterpreted term. It refers to the most profound scientific discovery of the last century. The relevance for philosophy is obviously 'realism'. The relevance for physics starts with that Einstein-locality (~ the light velocity limit) is questioned by many who misinterpret non-locality. The deeper relevance is a unification of the relativities, Einstein's and Everett's. In 2007, relational quantum mechanics already resolved the Einstein-Podolsky-Rosen (EPR) paradox. Those authors also implied that Everett relativity is not a mere interpretation. The situation now is similar to the way in which the twin paradox of special relativity was resolved. We undergo the same transition, this time toward an acceptance of Einstein-Everett relativity as deriving from a fundamental 'modal relativity'. Just like Minkowski diagrams clarify the twin paradox better than prose and equations, such issues are best addressed by visually intuitive models, because the terminology is grounded; words like "age" or "actualization" are less easily misinterpreted if they refer to features in a model, like lengths or areas that we can point to. I construct a short series of models as pedagogically as I could. They illuminate visually intuitively several otherwise almost impenetrable distinctions, such as that between three meanings of "non-locality". They show that single-future-*in*determinism is a pre-quantum concept. They show that the core of quantum mechanics is Peres' (1978)[1] more-disciplined-than-classical-common-cause correlations between *alternative* observations. All this becomes natural, as 'relative actualization' (~ "parallel futures") is natural in a 'modal paradigm' (Vongehr 2012a)[2]. The transitioning between paradigms in the proper Kuhnian sense (Kuhn 1970)[3], *requires* to forego much further general introduction by traditional 'language games' before 'showing'; in Wittgenstein's words, and he meant models (Wittgenstein 1958)[4].

# 2 The Non-Quantum EPR Many-Worlds Model

Einstein-locality, also called 'relativistic micro causality', implies that no influence travels faster than light. The limit is fundamental, as I defend in Section 6.4. However, "quantum non-locality", which is better called differently (Section 2.3), has been proven empirically (by experiment): If you assume a single "actual world" which "lives through time", meaning that it is not a somehow given space-time where the single future is already completely fixed, then certain hidden information would have to travel *faster* than light. The empirical evidence for this includes the experiments around the Einstein-Podolsky-Rosen (EPR) paradox (Einstein 1935)[5], because they violate the Bell inequality (Bell 1964)[6], which is explained pedagogically in the same notation as here in (Vongehr 2013a, 2012b)[7,8], where sophistries such as "superdeterminism" and photons conspiring to exploit loopholes are sufficiently rejected. Understanding what quantum non-locality actually refers to is central to understanding quantum mechanics. Therefore, my construction of the series of models starts with a simple non-quantum model that already has the non-locality which even many physicists fail to distinguish from true quantum non-locality, which we will recover later (Section 3).

**2.1   The EPR setup initially with Parallel Crystals**

The most accessible version of the EPR setup involves a source of pairs of photons. We will think of photons as being particles of light. Especially initially, we ignore that photons are observer-dependent quantum interactions. The source of photons is at Carl's place, who resides at $x_{Carl} = 0$ in the middle of our coordinate reference system (Figure 1a). Since we start with a non-quantum model, we may now imagine photons to be small mechanical systems. The two photons of each pair are separated by sending one along the *x*-axis to Alice and the other one into the opposite direction to Bob. Alice and Bob are both stationary relative to Carl, and both reside far away to the left at $x_{Alice} < 0$ and right at $x_{Bob} = -x_{Alice}$, respectively.

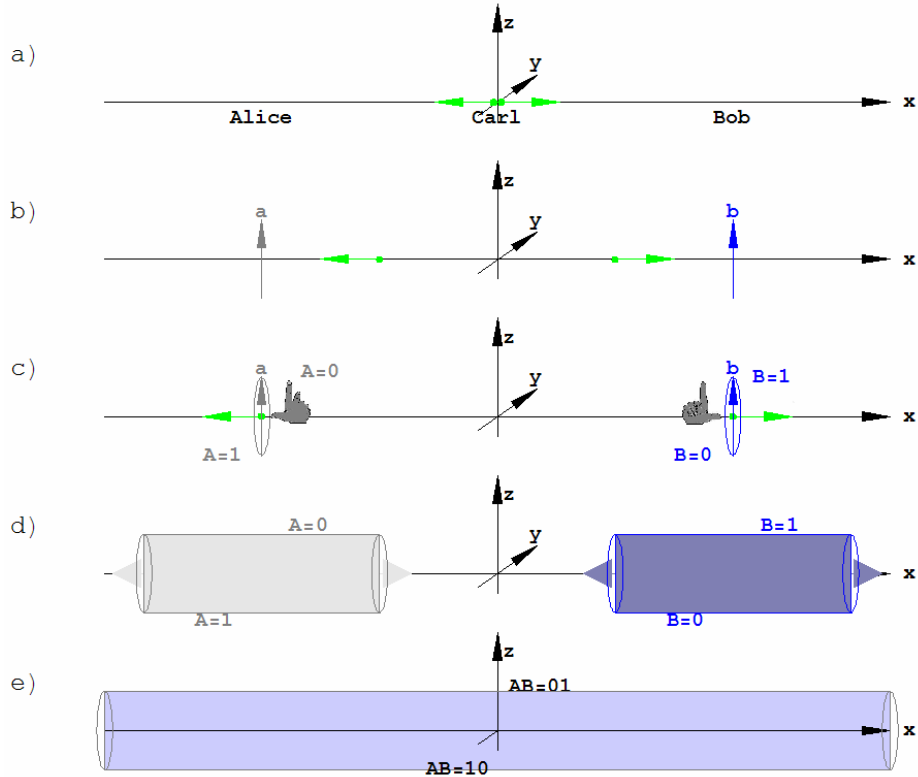

**Figure 1** Non-quantum many-worlds model: After a photon pair (green) is created (a) and separated (a to b), measurement directions **a** and **b** are selected randomly, but we initially consider only the parallel case (b). The photons arrive (c) and the measurement outcomes are represented by arranging them according to a right hand rule. This branching into different outcome 'worlds' propagates at the speed of light (d). If the measurement axes are parallel, two parallel worlds result (e), namely (*AB*) being (10) and (01). Both worlds may feel to be the only one and it can thus seem (under further assumptions) as if Alice's and Bob's measurements have influenced each other instantaneously (see Fig. 3). This is 'apparent non-locality'.

Alice and Bob both have a polarizing beam splitter (PBS), which may contain a calcite crystal for example. The PBS splits incoming light into two output channels, which both output linearly polarized light. This is similar to Polaroid sunglasses, the only difference being that the sunglasses absorb the photons of one of the output channels. Each output channel has a photon detector attached. Every single one of a number of experiments starts with Carl preparing a pair of photons. When the two photons are about halfway (more or less, that does not matter) on their paths to the crystals, Alice and Bob both rotate the crystals by random angles around the *x*-axis. Hence, when the photon pair was prepared, the angles were unknown. Since photons travel with the velocity of light, Bob's angle can neither influence Alice's angle nor her photon measurement outcome before long *after* the measurements have taken place. Alice's and Bob's rotations adjust the orientations of their PBSs' *z*-axes, which are **a** (= "$\vec{a}$") for Alice and **b** for Bob; see Figure 1b. It is not important how the *z*-axis of a PBS is related to its crystal's so called 'optical axis'; the reader does not need to understand what an optical axis is. A crystal's optical axis has no intrinsic negative-to-positive direction; a rotation of 180 degrees can lead to the same physical situation. This does not matter, because we are not here to dream up mechanisms about how fictional non-quantum photons may interact with fictional physics of crystals. I desire to model a causal event structure, which needs me to arrange the different possible outcomes of measurements according to suitable conventions so that the *necessary* symmetries can be embedded! The negative-to-positive direction along the PBS *z*-axis is a mere convention. Our conventions here must be modified later when modeling arbitrary angles, such as the anti-parallel situation where **a** equals the negative of **b** (meaning that they point in opposite directions). However, in order to arrive at 'apparent non-locality' with as few assumptions as possible, this section models only the case when Alice and Bob happen to adjust **a** and **b** in parallel.

   Alice receives only one photon of each photon pair. In one of two possible futures, Alice will detect that the photon exited the PBS channel that leaves light *vertically* polarized, meaning that the electrical field vectors of this light point along **a**. Another possible future-Alice observes the photon having left the other PBS channel, which leaves the output in *horizontal* polarization. The two channels are labeled as follows: since the vertical (or horizontal) is parallel (or orthogonal) to the PBS *z*-axis by convention, the measurement is recorded as $A = 0$ (or 1, respectively). These conventions are convenient because a so called 'cross product' is proportional to the Sine of the angle between the output polarization direction and the PBS *z*-axis. That Sine is here zero or one, respectively, and this also allows lay persons to use a simple right hand rule. Bob records $B = 0$ or 1 in the same way, and the reader should check now that Figure 1c correctly represents all the different possible measurement outcomes by arranging them according to that right hand rule: consider the arriving photon's propagation direction **p** (green arrow) along the thumb, and the *z*-axis of the PBS (grey or blue) along the index finger; then the cross product [which is thus $(-\mathbf{x}) \times \mathbf{a}$ and $\mathbf{x} \times \mathbf{b}$ for Alice and Bob respectively] points along the middle finger toward where we represent the measurement outcome 0; the opposite direction points to outcome 1 instead. These different outcomes are also called "branches", one branch for every possible outcome. Since I eventually model arbitrary PBS alignments, I already arrange outcomes on a circle around the *x*-axis. Each circle is split into two halves (Figure 1c), representing two possible outcomes.

The outcome of a measurement can be transmitted fastest via light signals. Thus, imagine the split circles to grow into split cylinders along the *x*-axis, away from the measurement events, and with the velocity of light (Figure 1d), thus broadcasting the measurement outcome, and modeling where the outcome can already be known. Loosely put: the "dislocations of decoherence" that travel outwards from the measurement events "split" (discussed in 2.4) the world like a "zipper" (Zeh 2010)[9] into the different possible outcome worlds. Where the cylinders have not arrived yet, a measurement cannot have any effect. The branching/splitting due to Alice's measurement has not arrived at Bob's place yet. The cylinders meet first at Carl's place. When that happens, the $A = 0$ outcome "meets" (discussed in 2.3) the $B = 1$ outcome, resulting in a $(AB) = (01)$ world, and $A = 1$ meets $B = 0$ to result in the $(AB) = (10)$ world. Finally, one cylinder split into two results or "worlds" (Figure 1e). Generally, there are *four* possible compound measurements $(AB)$ for every photon pair: (00), (01), (10), or (11). However, Carl prepares every photon pair in such a way (in the quantum case called 'singlet state') that *if the PBS are parallel*, only the measurements (01) and (10), short (U) for "Unequal", can result. This constraint is called anti-correlation: When **a** and **b** are parallel, (00) or (11), short (E) for "Equal", occur very rarely, and only because of slight crystal misalignments and other, for us irrelevant experimental noise.

### 2.2   Absolute Actualization and CCCC

The situation of you reading this section is now 'actual', meaning that it is not just a future possibility anymore, which it was a minute ago, but you 'actually' read it, experience the reading, being conscious of doing so. 'Absolute actualization' is the concept of that there is just a *single* world of outcomes actual at any one time (Figure 2a). The outcomes are not just actualized *relative* to us finding ourselves with that outcome, but according to absolute actualization, there is only this one world now. Absolute actualization insists on that only a single of all possible futures is real: "There are different possibilities, but there will be only one *actual* outcome, and the observers of that outcome are the only real ones, while observers of the alternative outcomes will not exist, not even in their own worlds, because I only accept a single world at any one time."

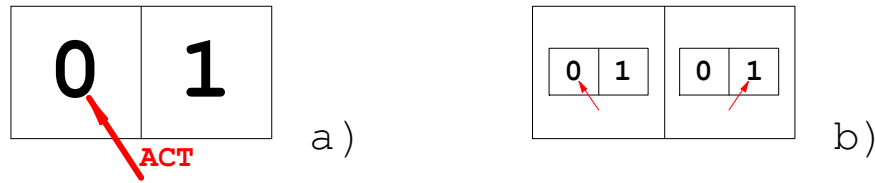

**Figure 2** a) *Absolute* actualization (red arrow "**ACT**") picks a *single* possibility and is equivalent to a model-external choice. b) Actualization is *relative* (multiple red arrows) in a larger model, one which includes the external of the smaller one, *including the alternative outcome*. I could write "including the alternative outcome's actualization", which shows that actualization is a mere regress step if it does not describe something, such as for example the act of observation by Alice. Thus, *absolute* actualization is itself *relative* to a model (or at best, relative to an observer) and absent from fundamental models of totality.

Absolute actualization can be refused "more philosophically" (Section 6.2), but you may reject that, and therefore, we may treat Figure 2b for now as if it fails to grasp the meaning of "absolute". In order to go on and see absolute actualization fail in models that describe empirical physics, the following is crucially important, especially in Section 3.4:

I) Absolute actualization assumes that already at the measurement events in Figure 1c, only *a single* Alice and only *one single* Bob are actual, …

II) … but they are *both* already actual! *Their lives are not put on hold* until after a single Carl possibility is determined later (Figure 1e). Neither is it in the spirit of absolute actualization of *single* outcomes if first *multiple* copies of Bob are actualized (~ alive, consciously observing) and then those inconsistent with Alice's measurement are killed and removed from history. Thus, absolute actualization must *already in this non-quantum model* assume either 'Classical Common Cause Correlation' (CCCC) or violation of Einstein-locality, as I will now describe them in detail:

**CCCC** means that both outcomes, Alice's and Bob's, are predetermined at Carl's photon source in the middle, and this information is carried by each photon. To model this, some direction orthogonal to the *x*-axis is somehow already singled out by the photon pair production event like in Figure 3a, represented by a pointer **ACT** (bold type like **a**, because **ACT** also has a direction), the "actualization pointer" (again in red). **ACT**'s angle in the $y \times z$-plane is carried by each photon (Figure 3b) to the measurement devices. This is the concept of 'hidden variables', which is similar to Bob getting the left sock of a pair of socks if Alice got the right sock, because this "anti-correlation" is based on the pair of socks as the *classical common cause*.

**Violation of Einstein-Locality:** Already in this non-quantum model, if there is no CCCC, the measurements must influence each other instantaneously, as illustrated in Figure 3c. **ACT** is perhaps decided by Alice's measurement and immediately known by Bob's photon. Such is *compatible* with the relativity of reference frames (because "spooky" superluminal interaction conceivably takes its velocity and causal order relative to Bob's and Alice's PBS crystals' combined center of mass reference frame), but *in*compatible with Einstein-locality.

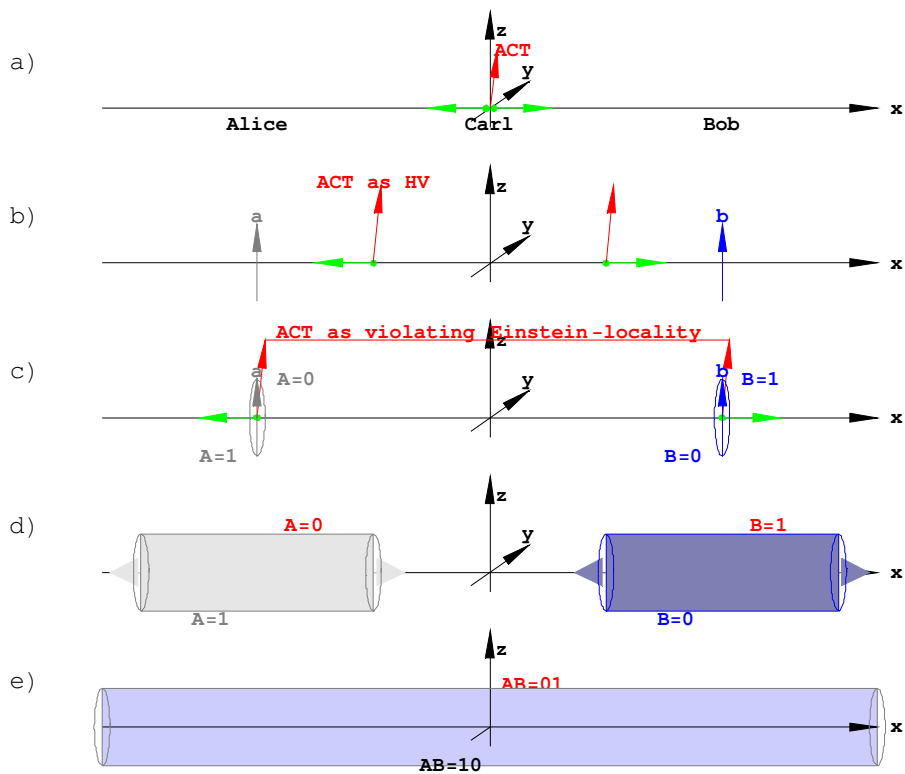

**Figure 3** Absolute actualization is modeled by a pointer '**ACT**' orthogonal to the *x*-axis. **ACT** can be a hidden variable (a, b), which is here still just an angle. Alternatively, the single random direction has been thought to be determined during the measurements as in (c), which would violate Einstein-locality. **ACT** is added to the model of Fig. 1 but makes no essential difference except for labeling one world "actual", here by printing preferred outcomes in red (d, e).

## 2.3 Three Different Concepts of Non-Locality

As far as the many-worlds model in Figure 1 is concerned, **ACT** does not modify it in any essential way; **ACT** only labels one world, painting it red perhaps (Figure 3d, e). From the point of view of a believer into absolute actualization on the other hand, all not actualized worlds are dead and the many-worlds model is a mere mathematical construct. As far as discussed, both interpretations may seem possible. A theory of everything (that is possible) must by definition account for every possibility somehow. Therefore, I take the concept of spaces where different possibilities stay arranged (for example the Hilbert space of quantum mechanics) seriously already during pre-quantum modeling. The many-worlds model correctly represents how different Alice-worlds and Bob-worlds stay arranged relative to each other in that mathematical space where possibilities are so arranged. [This does not imply a "belief" in some "real" branching space-time (BST) where space-time curvature destroys the overlap of different worlds, which, by the way, models decoherence due to gravity. A "real" BST cannot embed the Hardy paradox as far as I can see!] The model keeps the $A = 1$ and $B = 0$ outcomes correctly arranged so that they result in the (10) world. The model thus shows:

    1) How non-locality *appears* without violating Einstein-locality, therefore: 'apparent non-locality'.

    2) This model is not quantum, because CCCC and hidden variables (**ACT**) in the photons would still be sufficient to account for the statistical correlations. This removes the common misconception that "many-worlds" is automatically "quantum".

    3) Apparent non-locality resulted here from classical (anti-)correlation, which the EPR setup preserves *at all angles* as the classical contribution to the statistical correlation (Alice's input photon behaves *at all angles as if* orthogonal to Bob's output photon[7]). In this sense, apparent non-locality is precisely the *non*-quantum contribution to the statistical correlations. The contribution that is essentially quantum, which we will model in later sections, should not be related too closely to "non-locality", and the term "*Quantum* non-locality" for those "hidden influences" is misleading, because they only prove the insufficiency of ('locally real') hidden variables, but also non-quantum models may lack such hidden variables! For example, a model can conceivably have "photons" that lead to anti-correlated zero-or-one measurements but without the photons having

spin (or any other "hair"). That would equally imply a breakdown of Einstein-locality *only under the assumption of a single outcome world*.

Note that we must distinguish three different "non-localities":

    1) Violation of Einstein-locality (~ faster than light)

    2) *Apparent* non-locality compatible with CCCC

    3) A further quantum contribution toward the statistical correlations, which some experts refer to as 'hidden influences' rather than "quantum non-locality". This forces me to immediately add remarks that may be best only skimmed by lay persons:

### 2.3.1 Two remarks on "QM non-locality" versus "hidden influences"

"Hidden influences" sounds mystical and reminds of 'hidden variables'. However, the hidden influences are formally identified by the statisticians' formalism: They are statistical correlations between potential measurement outcomes; that a statistician seldom calls those outcomes "parallel worlds" just shows that there is no mystics here whatsoever. Alternative outcomes to any outcome that an observer observes are self evidently "hidden"; they are by definition not the observed outcome. The statistical correlations, the "influences", are between unobserved situations and thus hidden. In this well understood sense, "parallel worlds" are truly local hidden variables (LHV); they are just not 'locally *real*' hidden variables, where "real" means 'counterfactual definiteness' (Section 6.6).

Some may justify the term "QM non-locality" by holding quantum logic to be prior to emergent space-time. Einstein-Everett-relativity is then due to that instantaneous fundamental quantum interactions will appear as a light-cone structure in space-time. However, this light-cone structure is precisely what is called "Einstein-local", also historically long before quantum physics was known to be closely related. This non-locality along the light-cone is precisely what "*Einstein*-local" means. The term "*quantum* non-locality" is under this consideration even worse, and indeed, nobody writes "*quantum* micro causality" instead of "*relativistic* micro causality" either.

## 2.4 Introducing Probabilities, Cutting a Sausage, Probability *C*

> 'what we discover at work here is the round-robin of using "random" to arrive at the meaning of equiprobability, and equiprobability to arrive at the meaning of "random"' (Signorile 1989)[10]

I did not need to mention probability yet! To introduce it, there are several advantages in modeling by literally cutting a sausage. The final cylinder (the sausage) may be given from the beginning, just not cut yet. It is then cut by the measurements (Figure 4a). Sections of the cylinders' circumference (the sausage skin) can be identified by their angle relative to Carl's *z*-axis. The cylinder is not inside the coordinate space, but its visual representation is drawn around the *x*-axis because the angles matter. Different sections of the cylinder circumference (and later also of the cross section area) represent different measurement outcomes (worlds). When an un-polarized photon interacts with a PBS, the two different outcomes are equally expected. This equivalence reflects the symmetry of the setup. It does not follow from that the outcomes are equally probable. Probability as a measure of expectation instead follows from symmetry. Probability is thus grounded in the indifference toward any particular result if the possible results are equivalent due to symmetry. Throughout this work, I avoid Signorile's "round-robin" like this. We conserve the symmetry in our model by letting every outcome occupy half of the complete circumference each (for *A* on the left and similar for *B* at Bob's place). The length *C* of the arc on the circumference that a particular world takes up represents the classical (meaning 'non-quantum') probability of that world; remember this "*C*" to stand for "classical" as well as "circumference". The total probability is 100%, which is reflected by the cylinder's total circumference being unity; it's radius thus equals $1/(2\pi)$. The axes **a** and **b** cut the cylinder like cheese wires into two halves from both ends simultaneously (see black planes in Figure 4a). If **a** and **b** are parallel, they cut along the same plane and only two kinds of parallel worlds result (Figure 4b): one half measured $(AB) = (01)$, the other $(10)$, and both their classical probabilities are $C = ½$.

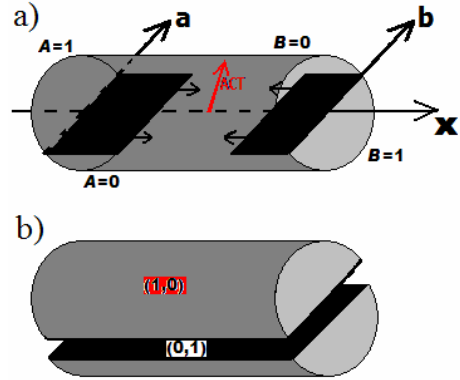

**Figure 4** The previous model illustrated by literally cutting: (a) The measurement directions **a** and **b** cut like wires and represent the decoherence that propagates at the speed of light. **ACT** is shown in red again. (b) If the measurement axes are parallel, two parallel worlds result, namely (*AB*) = (10) and (01), both having classical probability *C* = 1/2. In both worlds it seems as if the ends of the sausage knew instantaneously about the respective other end's measurement outcome. This 'apparent non-locality' is still not the essential EPR quantum correlation. **ACT** labels (*AB*) = (10) in this example.

## 2.5 Non-Quantum Many-Worlds with Arbitrary Angles

Carl measures the PBS crystals' *z*-axis angles as $\alpha$ and $\beta$ (rotation around the *x*-axis, relative to Carl's *z*-axis). For example, Bob could rotate his crystal to let $\beta$ be either 0° or 45°. More formal: $\beta = b\,\pi/8$ with $b \in \{0, 2\}$. Alice adjusts her crystal to $\alpha$ being either 0° or 67.5° ($\alpha = a\,\pi/8$ with $a \in \{0, 3\}$). Each has only two angles to choose from; this simplifies experiment and discussions. For instance, Alice can decide her angle choice by sending a 45° polarized photon (relative to her PBS) through her own PBS. Due to symmetry, this auxiliary photon is also equally expected to go through the horizontal or vertical output channels; none of these outcomes is preferred over the other. This procedure avoids certain "free will" discussions. Note that the models will later work at arbitrary angles, but to judge whether any model of the EPR setup reproduces quantum physics in an essential way, one should focus on only these few angles and insist on that the model violates the Bell inequality (explained below) instead of the so called CHSH inequality and so on; this has been defended[7] and is known as the 'Quantum Randi Challenge' methodology (Vongehr 2012b, Gill 2013)[8,11]. The relative angle between the crystals is then $\delta = (b - a)\,\pi/8$, merely four values. Testing for Bell violation needs only these four values of $d = |b - a| \in \{0, 1, 2, 3\}$, including $d = 0$, that is when **a** and **b** are parallel and the model in the previous sections (up to and including Section 2.4) applies. Calculating the Bell inequality itself only needs the three $d \in \{1, 2, 3\}$, which correspond to $\delta$ being the three so called 'Bell angles', which violate the Bell inequality most severely; see also the blue bars in Figure 6.

Due to the symmetries of the setup (including PBS crystals and singlet state anti-correlation), all four outcomes (*AB*) are equally expected at $\delta = 45º$. Therefore, we must modify the previous model and let a PBS crystal's internal *y*-direction cut the sausage just like its *z*-direction does, so that every measurement will split the cylinder/sausage like a cross of cheese wires into four equal pieces, all having $C = ¼$ (Figure 5a). The right hand rule (cross product) still applies and points toward where we start going counterclockwise on the circumference in order to represent the measurement outcome zero, meaning 0-worlds are in the first quadrant (for Alice's "cheese wire wheel", the first quadrant is wedged between her crystal's positive **z** = **a** and **y** directions) and third quadrant; these are those quadrants with black lines. 1-worlds are represented in the second and fourth quadrants, which are marked by green lines (for both, Alice and Bob; Bob's wheel is seen from the back in Figure 5b). The cuts again travel away from the measurement events; both crosses severing the whole cylinder like two pairs of crossed cheese wires slicing the sausage from both ends going inward. The crosses pass through each other, admittedly unlike cheese wires, in the middle at Carl's place, and thus the cylinder eventually falls into eight stripes with wedge-like cross sections, all literally *parallel* along the *x*-axis. Opposite wedges represent the same (*AB*); the physical symmetries force this complication into the model. Thus, the four different kinds of parallel worlds (*AB*) result, and all four are automatically colored differently in Figure 5b, for example, (11) has only green lines.

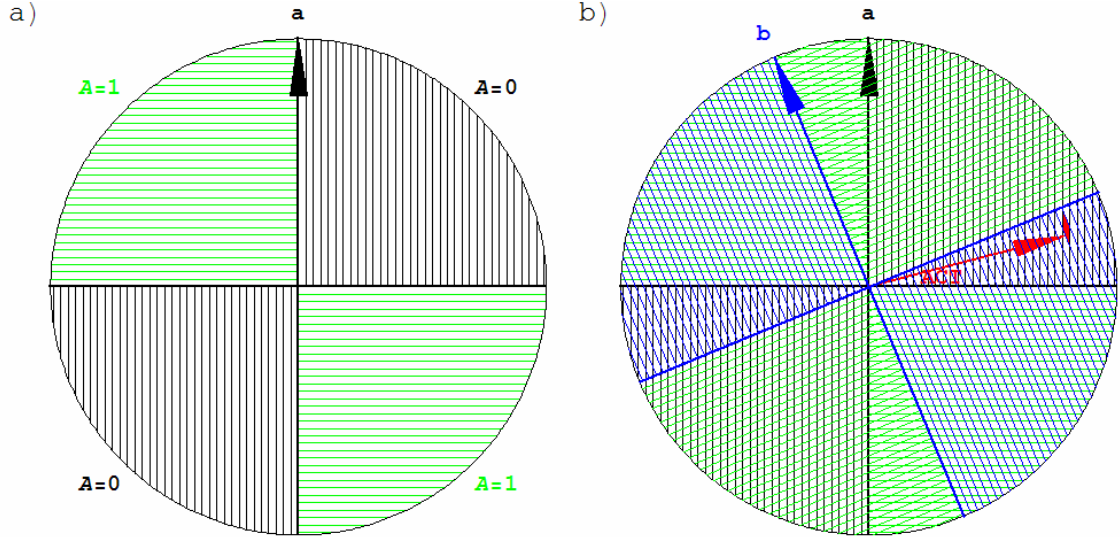

**Figure 5** The cross-section of the probability cylinder (as seen from Bob's end): a) Alice's measurement (black/green "cheese wire wheel") splits it into four quadrants, distinguishing two types of worlds by *A*. The lines inside the quadrants are parallel to the output photon's E-field vectors. b) When both measurements' decoherences overlap (Bob's in blue/green), the cylinder has branched into four types of parallel worlds (*AB*). Their circumference sections' lengths (as well as their wedge-shaped areas) equal *C(AB)*, which depends on *δ* (= apparent non-locality). It can still not correspond to the quantum probability *P*; **ACT** still points with classical probability *C*. *P* depends on the squares of state vector projections, being thus related to dot products between the measurement directions. Hence, the number of diamond areas like the red one which **ACT** points to, if identifiable with distinct worlds, almost reproduces quantum mechanics.

For $|\delta| \leq \pi/2$, the classical probability *C(AB)* of the parallel worlds are

$$C(\mathrm{E}) = 2|\delta|/\pi \quad (= d/4), \qquad C(\mathrm{U}) = 1 - 2|\delta|/\pi \quad (= 1 - d/4) \tag{1}$$

, where "E" and "U" signify "equal" (*A = B*) and "unequal". *C* is thus the straight line in Figure 6. *C* depends on the relative angle *δ*, which is only known wherever both cuttings, Alice's and Bob's, arrived already; not at the measurement events. Therefore, the model can be said to have further apparent non-locality beyond the mere anti-correlation. However, creating a single hidden variable together with the photon pair, namely the direction **ACT**, proves that this model still allows hidden variables. The apparent non-locality can still be assured by CCCC instead of only by the many-worlds structure alone.

### 2.6 QM-to-Classical Consistency, QM Probability *P*, Bell Violation

The angle dependence of the correct probability (as observed in nature) is expected from the usual behavior of light at polarization filters. The classical physics involves a projection of the incoming electric field vector onto the PBS axis. If the angle between a PBS *z*-axis and the polarization of the incoming light is $\gamma$, the projections' length on the z-axis is cos($\gamma$), that along the PBS *y*-axis is sin($\gamma$); in a fully pedagogical presentation, these trigonometric functions can be replaced by talking only about projections (casting shadows). Energy is conserved at all angles $\gamma$, which is consistent with that energy depends on the square of the electromagnetic field vectors of light: the outgoing light intensities in the two channels are thus proportional to $\cos^2(\gamma)$ and $\sin^2(\gamma)$, which always sum to 1 (Pythagorean theorem). These factors must therefore also be a photon's probabilities for appearing in the output channels 0 and 1, because energy is directly proportional to the number of photons! Careful: This does not change the fact of that the photons in the EPR experiment are locally (at Alice's and Bob's place separately) expected (and observed) to equally likely appear in each of the two PBS output channels, distributing the incoming photons and thus their total energy usually evenly over the channels. However, if Alice's measurements are later compared with Bob's, it must be that Alice's *input* photon behaved ***as if*** (in more detail explained in [7]) orthogonally polarized to Bob's *output* photon – for every pair Carl prepared. In other words, quantum-to-classical consistency, here argued by the energy conservation of the classical description of light and the photon description together, demands that the compound measurements' probabilities are $P(01) \propto (\vec{a} \bullet \vec{b})^2 = \cos^2(\delta)$ etc., or generally:

$$P(\text{E}) = \sin^2(\delta), \qquad P(\text{U}) = \cos^2(\delta). \qquad (2)$$

The Bell inequality in terms of measurement counts $N_d(AB)$ is the relation $N_1(\text{U}) \leq N_2(\text{E}) + N_3(\text{U})$. Quantum contributions to probability are those which violate the Bell inequality. *P* lets us expect that the inequality is violated: $P_{d=1}(\text{U})$ alone is *larger* than $P_{d=2}(\text{E}) + P_{d=3}(\text{U})$; see blue bars in Figure 6. Therefore, *P* is a quantum probability, because it does something which the classical *C* cannot do. The blue bars make it obvious that the violation of the Bell inequality is directly correlated to the bulging away (from the straight line *C*) of the probability curves in Figure 6. *C*, being the straight line itself, cannot violate the Bell inequality [check it: (1 – 1/4) = (1/2) + (1 – 3/4)]. However, the model does as good as a non-quantum model can do, namely reaching equality between the left and right hand sides of the Bell inequality ('saturating' it). *C* is also in this sense *the* classical probability in the modeling of the EPR setup!

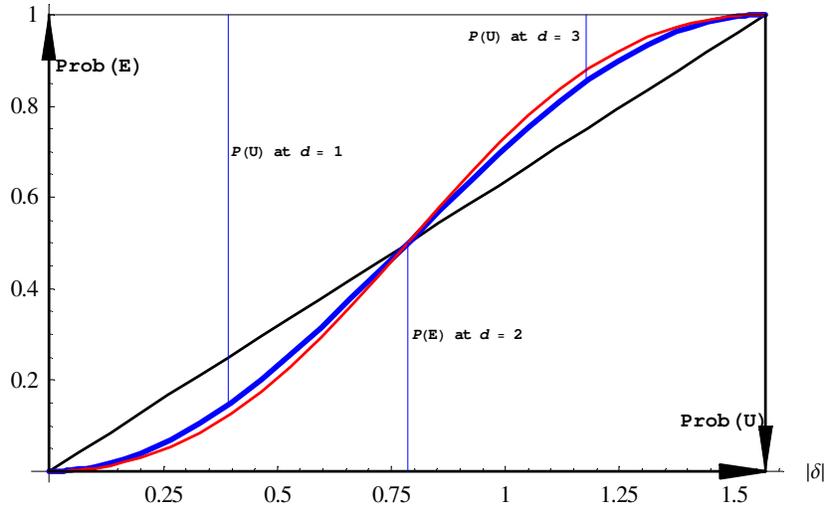

**Figure 6** The probability of (*AB*) being equal ("E") and unequal ("U") (right hand axis going down), plotted for $\delta$ between zero and $\pi/2$. All models agree at multiples of $\pi/4$. The non-quantum model's *C* gives a straight line (black diagonal). The correct *P* of standard quantum mechanics results in the thick blue curve bulging away from the straight line. The three terms of the Bell inequality are presented as blue bars for *P*. The bulging corresponds directly to violating the Bell inequality. The *P*\* of our transition model (see below) bulges even further (red curve) and is thus a non-standard quantum model.

## 3 The Classical-to-Quantum Transition

The following distinction makes as of yet no difference in the resulting probabilities, which may initially help to accept that one can indeed think of randomness in two ways.

### 3.1 Model-*Ex*ternal Randomness

If we select the direction of **ACT** at random, for example by twirling a pencil on the center of Figure 5b, it will point with probability *C*(11) to the (11) part of the circumference. The fairness (fair randomness) of the pencil twirl depends on that we, the modelers outside of the model, actually do it randomly. The randomness is *external* to the model. Nothing inside the model forces us to be fair. In fact, I put **ACT** where it fits best didactically in the figures, not randomly. Model-external randomness is closely related to absolute actualization. They both together are relative to the model (Figure 2). They are together modified if we nest a model into a more comprehensive model, one that models "the outside" that is external to the smaller model, something that has been omitted, like the omitted alternatives (Figure 2b). Signorile's "round-robin" charge of circularity rather than regress (Section 6.5.1) misses that the vital issue is the nesting of models in more general models. Moreover, since external randomness is thus not the same as the concept of 'objective chance' (whatever such is supposed to be), the external/internal distinction is not the 'objective vs. subjective' one (Section 6.5).

### 3.2 Model-*In*ternal Randomness and Angle Communicating DOF

The compound measurement requires not just observing Alice's and Bob's results. In order to infer the dependence on the relative angle, the angles must also be communicated. Alice and Bob may float in space and send many particles so that Carl can deduce the angles. Or: the whole EPR setup is fixed to one huge laboratory floor and Alice (and similarly Bob) transmits only a single photon whenever $a$ (or $b$) is 0, while the laboratory effectively communicates the crucial information which encodes the physical $\delta$. Regardless, there are always many degrees of freedom (DOF) involved, namely some sort of microstates which distinguish between different $\delta$, because we could test at as many different $\delta$ as our experimental means can distinguish, not just the four $\delta$ we mainly discuss. Think of the circumference (in Figure 5a or b) as divided into many small sections, each representing a different world distinguished by those 'angle-communicating DOF' which are entangled with the measured photon pair under consideration. Model-internal randomness arises from that all the Alices in all these parallel worlds do not remember any bias in the expectation she had toward which world she would find herself in, even if she knew all the details of the DOF of every possible future world. Those future worlds have been equivalently expected possibilities, equivalent *by symmetry*, just like all the different possible sequences of ten flips of a symmetrical coin; nothing here is equivalent due to probability functions which we did not construct yet! I again avoided Signorile's round-robin. Model-internal probability $Pr$ derives from that symmetry, from that a symmetric coin coming up tails is fundamentally "just as irrelevant" as it coming up heads. If among all futures that are equivalent according to their angle communicating DOF, those that are (U)-worlds outnumber (E)- worlds, Carl will be the more surprised in an (E)-world the fewer of those there are compared to the U-worlds, in precisely the same way we feel surprised if we witness a symmetric coin coming up tails ten times in a row. The model-internal probability for (E), written $Pr$(E), is a measure of rational expectation (Deutsch 1999)[12] or even better of evolved surprise, and is thus well defined if equal to the number of E outcome worlds $N$(E) divided by the total number:

$$N(\text{E})/[N(\text{E})+N(\text{U})] \quad \in [0,1] \tag{3}$$

The closer $Pr$(E) is to zero, the more surprised Carl will be to find himself in an E-world, also because at $N$(E) = 0 (such as if $d$ = 0), there is no such world, so how could he find himself in it. If $Pr$(E) = 1, he is not surprised at all because there are no U-worlds.

### *3.2.1   Model internal Memory of Sequential Measurements*

After many runs, in the majority of worlds, the statistical frequencies recorded in single worlds reflect automatically $Pr$ closely; this can be clarified with small numbers: Assume three new branches, $N$(E) = 1 and $N$(U) = 2, after a single photon pair. After $i$ = 3 photon pairs, there are $(1+2)^i$ = 27 worlds, only one of which observed the sequence (EEE), 8 are (UUU), and 6 worlds record two E versus one U like (EUE). The largest fraction of 12 worlds comprises 44% of the 27, and those 12 all contain a single E for every two U outcomes, like (UUE) or (UEU), precisely the ratio for a single run. The math-phobic should skip the following: Generally, the number $Q$ of outcome worlds with $r$ occurrences of E is:

$$Q = \frac{i!}{r!(i-r)!} N(\text{E})^r N(\text{U})^{i-r} \quad ; \sum_{r=0}^{i} Q = [N(\text{E})+N(\text{U})]^i \tag{4}$$

Adding for example $i/3$ values of $Q$ around the most common $r$, an ever larger fraction of worlds falls in such a not-surprising range: what was 44% for $i$ = 3, becomes 57% at $i$ = 6, and 70% for an experiment observing $i$ = 9 photon pairs. This is another way to see that $i$ = 800 is more than enough[8] for distinguishing classical from quantum models.

Graham (1973)[13] failed, here I agree with (Kent 1990)[14], to show that certain axiomatic models give rise to unsurprising statistical frequencies. Since then, such has been attempted and rejected often. The failure is usually that the numbers of worlds are not well derived even for single measurements. I support the correct ratios for a single compound measurement by classical-to-quantum consistency (2.6), assuming photons, which is a quantization demanded by further consistency arguments which do not involve probability laws. In any particular of my models, the *Pr* of a single run either equals the correct *P* or not. Sequential measurements are then trivial, because they automatically reproduce *Pr* as shown above.

### 3.3  Many Wires cutting Areas, Internal Randomness going Quantum

*If* the worlds on the circumference of the cylinder are all equally sized, the ratio $N(E)/N(U)$ equals $C(E)/C(U)$, and external and internal randomness give rise to the same statistics! The model-external and internal descriptions can stay consistent also if the *areas* (multiplied by $4\pi$ for normalization) in the cylinder's cross section serve instead as the probability measure, meaning for example that the whole wedges below the (11) arcs are taken to represent all the (11) worlds. The areas in these wedges sum to $C(11)$. Model-external randomness could now involve throwing a pebble onto Figure 5b in order to determine to which spot **ACT** points to – thus, **ACT** would now also carry length information and be no longer just an angle, but this is an unimportant detail. Again, if the worlds are further distinguished by many angle communicating degrees of freedom, now resulting in many sub-areas like the red diamond in Figure 5b perhaps, but *all of equal size*, counting the worlds will again result in the same probability as comparing the areas of the wedges, thus $Pr = C$. Since the pebble toss can be done in advance and its result conceivably be hidden in the photons, such models cannot violate the Bell inequality. They are all *non*-quantum!

However, now think of the many lines in Figure 5a as if they all cut the cylinder, just like the cheese wire cross before. The cheese wires are now only a tiny separation *s* apart so that there are many of them and every quadrant is cut into 1/s parts (by each cheese wire wheel, also Bob's again of course). The resulting worlds, each a diamond like the red one in Figure 5b, have each an area of $s^2/\cos(\delta)$ in U-wedges, but $s^2/\sin(|\delta|)$ in E-wedges (we ignore small irregular areas at the circumference for example; this modeling error is reduced with more cheese wires). The number of worlds $N(E)$ is therefore the wedge area $C(E)$ from Equation (1) divided by the size of the sub-areas:

$$N(\text{E}) = \frac{1}{s^2}\left(\frac{2|\delta|}{\pi}\right)\sin(|\delta|), \qquad N(\text{U}) = \frac{1}{s^2}\left(1 - \frac{2|\delta|}{\pi}\right)\cos(\delta) \qquad (5)$$

, for $|\delta| \leq \pi/2$. The model-internal probability *Pr* (Equation (3)) of this model, let us call it *P\**, violates the Bell inequality stronger than the quantum probabilities *P* (Figure 6). This model is thus a non-standard quantum model. It has another problem: The total number [*N*(E) + *N*(U)] is smallest at $\delta = 45°$, at least for sufficiently many cheese wires. Therefore, if we include the angle choices into the many-worlds modeling, for example by presenting Figure 5b for the four different *d* together in parallel as parallel worlds, most worlds contain empirical records where $d = 0$ outnumbers $d = 2$. We will fix this detail in a better model below. Important is that *P\** violates the Bell inequality and thus *P\** is essentially quantum. The model is therefore called *transition* model: it shows the *crucial step* from non-quantum to quantum physics, the step that *breaks the consistency between model-external and internal randomness* (here accomplished by U- and E-worlds having different sizes, but that is a mere technicality of the modeling). One key aspect that we will come back to is: **Since external randomness is like tossing a pebble, the external probability for (11) is always the same *C*(11), no matter how many worlds appear.** This can also be understood as follows (Figure 7): Absolute actualization means that Alice is "carried by the flow of time" (Section 6.5.1) along one unique path, her history, in the branching tree of outcomes, step by step. Once she gets into the (11) wedge with probability *C*(11), any branching of the wedge into sub-branches is no longer about the probability to observe (11) – that choice is made and done with. But there is no "flow of time", another model-external concept; Alice just finds herself in the world she happens to find herself in, and under this model-internal randomness, the number of sub-branches matters (to how surprised she is finding herself in any particular kind of them). Any particular history (green line in Figure 7) looks back through the tree and does not care about other branches. This supports the illusion of absolute actualization.

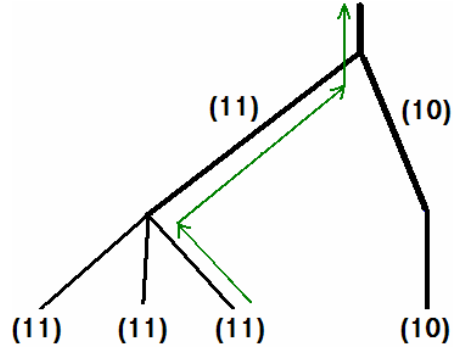

**Figure 7** Assuming absolute actualization proceeding through time, we must start for every trial again at the top of the tree (~ the next time the universe plays), thus the probability to observe the outcome (11) is ½. Model-internally, that probability is instead ¾. An empirical record (green line) looks backward and does not include other branches. The splitting of the single (11)-branch in this figure models the (11)-wedge being cut into many more pieces starting at Carl's place (*not* so called 'extra branching' beyond the available physical degrees of freedom).

*C* is a bad model for probability, because it is chained to an external viewpoint; it just happens to fit in case of non-quantum many-worlds models. That *P\** is so close to nature's *P* is irrelevant. Only because *P\** abandons model-external randomness can it improve over *C* at all.

### 3.4  The Crucial, Einstein-Local Step destroys Absolute Actualization

Because of Einstein-locality, the appearance of Bob-worlds (many Bobs) on the right cannot depend on Alice's to the left. Carl is in the middle, where the worlds from Alice- and Bob-measurements first meet; Carl is the first one able to observe the compound measurement. Therefore, naturally, more measurement outcomes appear at this point. At this event, the angle-communicating DOF become important. Combining those of Alice and Bob leads to many possible combinations; many more outcome worlds appear (Figure 5b). This accomplishes the crucial step, and it does so *locally* (starting at Carl's place, again broadcasting with light velocity). This single crucial step, namely the Einstein-local branching due to the total compound measurement (of *A*, *B*, *and δ*) accomplishes two aspects simultaneously, stressing that both, I and II are of the same essence, the essence that makes quantum physics quantum:

(I) It turns the model quantum: The consistency between model-internal and external randomness is broken so that *P\** can violate the Bell inequality. Since the model-internal $P^*$ depends on the number of worlds, we can even adjust the model by changing the sub-worlds' numbers. As soon as $N(\text{E})/N(\text{U}) = \sin^2(\delta)/\cos^2(\delta) = \tan^2(\delta)$, the internal probability $Pr(\text{E})$ of Equation (3) is $\sin^2(\delta)$, the standard quantum $P$ of Equation (2).

(II) **ACT** can no longer actualize: When choosing **ACT** while creating the photon pair, the axes **a** and **b** are not yet selected. **ACT** cannot be rigged to point with a probability $P$ that depends on a future alignment – this was proven by Bell, but we can now appreciate this intuitively as follows. **ACT**, after first having decided just for a wedge like $C(11)$ like in Figure 7, could with angle dependent statistics (post)select one of the new worlds when Carl observes. However: 1) **ACT** must at least already decide for a section like $C(11)$ before Carl can observe, because otherwise it cannot actualize a single Bob (say $B = 1$) and a single Alice (like $A = 1$) before – remember: their lives are not on hold until Carl observes. 2) Since the newly appearing worlds inside $C(11)$ are all (11) worlds, the probability does not change, which is the bold printed key statement in Section 3.3 again. The probability "to go from" the $C(11)$-cylinder section into a (11)-world rather than into a (10) or (01) world is precisely 100%. The probability stays to be $C(11)$ however much **ACT** may jump inside of a $C(11)$ wedge, however much you may like to deform the sub-areas according to $\delta$.

**ACT** represents the physics that is traditionally claimed to actualize a single future, pointing first to one situation, and then going to a single next one among multiple possibilities. **ACT**'s failure in the quantum models stresses that quantum mechanics implies *equivalently* expected futures. Quantum mechanics is not about some weird probability with which a single future among the possible ones is actualized, which would have to be a solipsist description. Quantum mechanics reflects the inclusion of all that is equivalently possible, equivalently into a theory of everything that is possible.

## 4   Improving the Quantum EPR Many-World Model

### 4.1   There is no Cylinder

Let us abandon a division of a given shape, cutting sausages. On encountering Bob's $B = 1$ worlds, the $A = 1$ worlds from Alice combine their angle-communicating DOF with them, resulting in many more combinations. Quantum-to-classical consistency demands a number proportional to $(\vec{a}\times\vec{b})^2$. Whatever details of the DOF may ensure this (further discussed in Section 6.3), crucial and thus relevant here for us is: this local branching due to the total compound measurement (of *A*, *B*, *and δ*) accomplishes the two aspects I and II together, now yet more obviously than already with *P*\* above:

(I) It turns the model into the *standard* quantum physical one: The model presented in Figure 8 models *δ* = 0 with *M* units put orthogonal to **a** (or **b**, that does not matter). For larger angles, take *m* steps away from that direction and walk them along **a** instead, thus (*M* – *m*)/*m* = tan(*δ*). The angles are therefore some form of quantized DOF and the resolution depends on how many different combinations they allow, as is usual and proper in quantum mechanics. Comparing especially the larger wedges in Figure 5b with the corresponding areas in Figure 8 reveals that this model is a modification. There are $2m^2$ (01)-worlds, $2(M-m)^2$ (00)-worlds, and so on, at every angle. Therefore, *N*(E)/*N*(U) equals $\tan^2(\delta)$, and the model-internal probability is the standard quantum *P*.

(II) **ACT** can no longer actualize: **ACT**'s pre-selecting of only a direction and then later selecting a small area along that direction fails as discussed before. If fully pre-selecting, **ACT** must choose symmetrically around the *x*-axis, because the axes **a** and **b** are not yet selected. However, worlds are no longer restricted onto a circularly symmetric shape. **ACT** will soon point to where there did not grow any world (see Figure 8). External flow of time and external randomness would switch the universe off by selecting no further world.

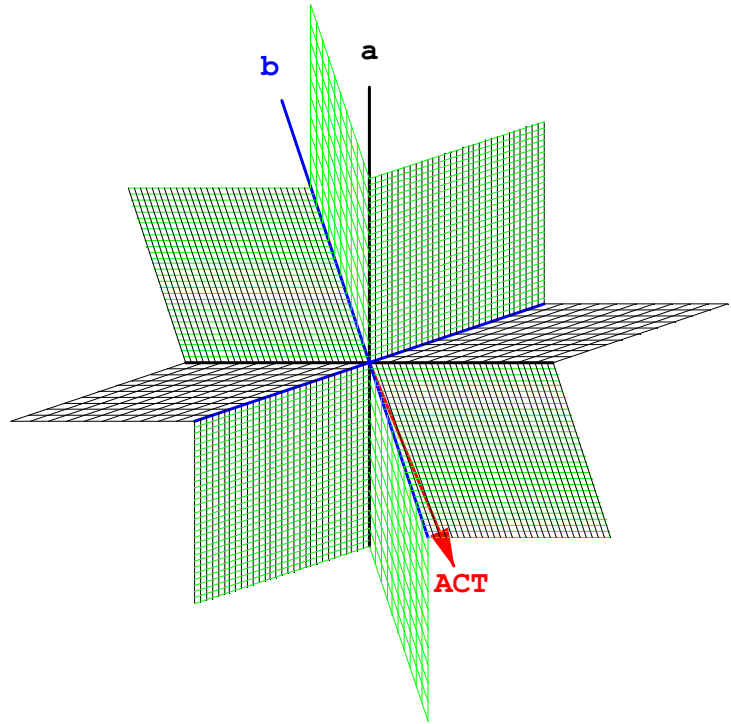

**Figure 8** A model that assumes so few degrees of freedom that single worlds are still discernable in the picture. The resolution is only $M = 40$ directions per quadrant; $m = 30$ determines the angle to be around $\tan(\delta) = 1/3$. (11)-worlds are purely green again, etc. This model contains the standard quantum probabilities. **ACT** tried to predetermine the 'actual future', but failed to hit any.

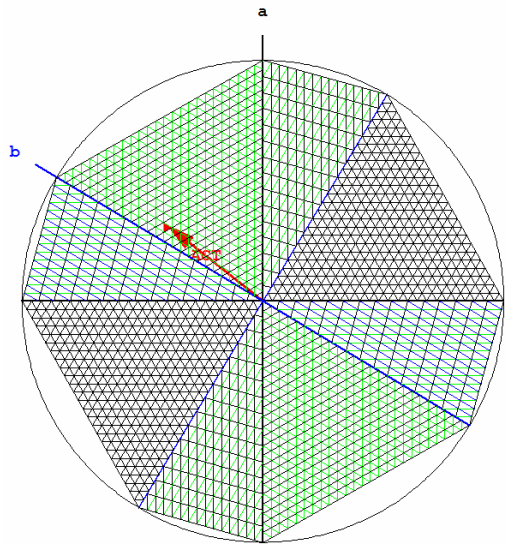

**Figure 9** Same model as in Fig. 8, but with different angle ($m = 15$), a different random position of **ACT**, and with the areas all put into triangles inside the original cylinder. The similarity to the transition model $P^*$ is obvious, clarifying that a deformation of its areas can indeed adjust the transition model's probability.

Figure 8 and 9 present the simplest model that recovers the standard quantum *P*. The total *N*(E) + *N*(U) equals 4[(*M* – *m*)² + *m*²]. This total can be fixed to be the same size for every *δ* by letting *M* depend on *m* and by increasing the resolution of the model by sufficiently large numbers. If that seems unsatisfactory, one can instead fix the total to $2^q$ by anticipating spin multiplicities of angle communicating DOF via binomial factors $N(E) = \sum_{k=0}^{p(d)} q!/[k!(q-k)!]$. However, it may be impossible to keep it visually presentable. The squares of *m* and (*M* – *m*) fit well into the two dimensions of Figure 8, but we ultimately model a high dimensional space (Hilbert space) by squeezing all its dimensions into the cross section orthogonal to the circumference (see discussion in Section 6.1). Binomial factors describe hyper-tetrahedrons, and depicting them squeezed down into two dimensions is not didactically promising. These models with standard quantum probability *P* are only presented for completeness. Relevant is the visual accessibility of the 'crucial step' in the *transition* model (Section 3.4).

## 5   Quantum Many-Worlds Kinetic Sculpture Science Exhibit

The here supplied models resolve the EPR paradox similar to (Smerlak 2007)[15], but have a didactically valuable twist: All are many-worlds models with apparent non-locality, but they are initially non-quantum. A single, natural modification turns the transition model quantum. This *local* modification (starting at Carl's place) obviously does not modify *locality*, but "*reality*", because it lets **ACT** fail. The models should find their way into undergraduate physics and philosophy lectures, but with more illustrations, much of this work can be understood by advanced high school students.

The transition model can be interpreted in two ways: 1) Traditionally, as a direct realism with absolute actualization (**ACT**) resulting in classical probability *C*, which can be practically experienced with a touchable exhibit made from two "cheese-wire-wheels" like in Figure 5a, mounted on an horizontal axis and, for example, red ping-pong balls instead of a single pointer **ACT**, one per student, bouncing and coming randomly to rest (before the wheels are adjusted according to two coin flips perhaps) inside a circular area on an exhibition hall's floor below the model's center (~ at Carl's place). The cheese wire wheels can be made of colored transparent stripes so that a pattern of parallel worlds

Figure 5b results. It can be projected onto Carl's circular ball-resting area. Larger groups of students can be guided to see for themselves how their ping-pong balls obey the Bell inequality, for example by putting all the balls that rest on (00) or (11) areas into a transparent tube labeled "Equal" so that (given a fixed total number of balls) the tube will be almost always filled up close to the straight diagonal line when the tube is held upright at the angle $δ$ in front of a display like Figure 6. 2) The model internal probability can then be gotten by counting the worlds in the projected pattern, thus closely reproducing the blue bars in Figure 6. It should then be briefly explained how Bell proved that no single world model can reproduce the experimentally found blue bars like the many-worlds model just did! Then, a few details about why the absolute actualization failed and cannot be saved should be provided. This is best done in a video animation of models similar to Figure 1 and Figure 4.

I claim that together with the obvious presence of apparent non-locality from the trivial, sausage cutting start, the transition model is so intuitive that it should be widely taught, including via this, first ever touchable and functioning quantum model exhibit.

## 6 Discussion/Supplemental Section

The discussion has many parts directed at experts in different fields. Skip as needed.

### 6.1 Re-explaining the Construction to Clarify the Transition Model's Nature

I started from a traditional viewpoint consistent with model-external randomness. Therefore, I modeled consistent with throwing a pebble onto sections of a homogeneous (to conserve fairness) 100% probability pie (all outcomes accounted for). Measurement angles are randomly chosen after the photon pair is produced. **ACT** (the pebble) cannot know the future orientation of a noncircular shape on which to pick a spot (Figure 8). The pebble can be assured to never miss the pie only if the pie has circular symmetry. Symmetries also fix $P$ to agree with the classical $C$ at $\delta$ being multiples of $\pi/4$. The strongest correlation possible via classical common cause is *complete* classical correlation, which leads to the straight line $C$ in Figure 6. However, quantum-to-classical consistency (Section 2.6) demands that measurements are correlated even stronger[1] (the bulging), namely according to probability $P$ as given by the thick curve in Figure 6. Less than complete classical correlation leads to a bulging into the wrong direction, which is the well known Bell proof[7]. $C$ is as close as one can classically approach to what we desire to model. After that, the transition model showed that in order to improve further, we need to abandon model-external randomness, which is chained to $C$. The transition model is a hybrid in the sense of that we still cut some unphysical 100% pie and get $C$, but we cut it into worlds that result model-*internally* in a different probability $P^*$. That the model still cuts a pie is a mere pedagogical device. It allows two very different interpretations of the same model, because sections of the circumference as well as the corresponding areas of the cross section both equal $C(AB)$. Cutting the area differently leads to $P^*$, but the $P^*$ model already should no longer be thought of as cutting anything already present rather than growing new branches. Here is why: The circumferences to the left and right each only model two types of worlds ($A = 0$ or 1 on Alice's side). In formal quantum theory, they are new branches that are orthogonal to each other in state-space. We indeed put them into a new space (not modifying what was there before, not cutting), namely the cylinder circumference. The branches are even orthogonal in Figure 5a; the green lines are orthogonal to the black ones. When Carl can then measure, again new branches have grown, naturally modeled by growing into the cross section, while the circumference (what was there before) stays again untouched. We must stay in three dimensions for visual intuition, thus, we cannot illustrate the new branches' being orthogonal to each other. However, a non-trivial growth of branches on top of the otherwise untouched circumference is in a sense precisely what happens in the more formal quantum mathematical description.

### 6.2 Relative Absolute vs. Verification Transcendent "Absolute Absoluteness"

"Actual" is always relative to possibilities, to possible worlds (~'modal relativity'), including Galilean temporal relativity as possible clock readings. Actual*ization* is only meaningful as a process relating different times: Some situation may not be actual yet, but actual in (~ 'indexical' to ~ *relative* to) a possible future. The impossible can by definition not be actualized. "Possible" therefore implies that we tacitly already model like Figure 2b. Figure 2a thus depicts a model-external choice related to 'model-external randomness' (3.1). *Absolute* actualization in one model is *relative* actualization in a more general one which includes the other possible actualizations. Thus, 'absolute actualization' is itself also only meaningful *relative to a model*. The absolute/relative dichotomy itself is relative to a domain of discourse, always model-dependent. For example, Einstein's *relativity*, even the general one, can hold in an *absolute* cosmological Einstein-ether background, which can be (but is not necessarily) *relativistic* in a certain more comprehensive description.

However, often the intended and claimed meaning of "absolute" is precisely that which is "absolutely" not relative. This "absolute absoluteness"-regress belongs to a different language game that aims for (not logically) 'consistent' *justification*, namely of the power that prohibits the relativization of some absolute, rather than *clarification*. In that language game, "model" means "*merely* a model", while "totality" is used as if referring to something that cannot be modeled. But language is a form of modeling. Uttering "totality" is modeling "totality", inherently pointing to some sub-structure held together by its relations with the rest of language, just like in a drawing. Those who hold on to the "absolutely absolute" (for example to "genuine stochasticity" while refusing the charge of committing a 'random randomness' regress that implies an external to "totality"), must, in order to be consistent, opine that modeling further is impossible and that any attempt to do so goes too far. Such rhetoric refuses to realize, let alone admit, that it itself went too far, that it said something that cannot be said (in the early Wittgenstein's words), and in this sense claims to model what cannot be modeled! I merely claim a self-consistent description. The possible fundamental description holds itself possible. I only aim for possible descriptions. In those, totality leaves by definition nothing external; a possible theory of everything (ToE) possible must fundamentally include all equivalently possible actualizations equivalently. Many-worlds are therefore tautological and pre-quantum as long as there is no argument for why possibilities correlate in Bell-violating ways.

### 6.3 What Microstates, what Parallel Worlds?

Statistical mechanics assesses the probability of a classical macrostate by the number $N$ of quantum microstates which can give rise to that macrostate. When modeling quantum measurements like (*AB*), the thing to be explained is already a quantum microstate. The microstates to be counted are now no longer found by looking further inside of that state. The angle communicating DOF could conceivably be further illuminated by putting together an EPR setup out of spin ½ particles, which provides multiplicities of spin alignments of all involved particles. Total angular momentum involves Clebsch-Gordan coefficients which are derived from spin ½ compositions with no bias for any of the spins to be either up or down. However, the photon pairs are prepared in singlet states; preparing other pair states results in a different outcome statistics. The sequences or spin alignments of other exchanged photons can therefore not be directly the relevant DOF. The microstates are ultimately in the macrostates of the measuring (recording) systems. This repels a common misconception. Bob is a recording system that we assign his own "existence" to *in just as much* as we assign "existence" to ourselves, whatever "existence" may mean. Every possible Alice accepts *in just as much* the "existence" of all possible Bobs, those that measured $B = 0$ and those that saw $B = 1$; nothing more is the "reality" of '(tautological) modal realism'. However, I do *not* accept the "existence" of many worlds with different photon polarizations before polarization measurements! Such would not solve the EPR paradox, because such **ACT** pointers, also if together in a large statistical ensemble, each and thus all fail as described. Considering spin ½ particles, many-worlds modeling does *not* claim two classes of worlds along every measurement direction attached to every particle, say on grounds of that there are two possible spin measurement outcomes along every direction one could measure. Worlds are perspectives of systems that have a perspective, of recording systems which physically involve decoherence. Somebody may have measured the spin in some direction and is about to tell me the direction and measurement outcome via photons; now those are very many worlds, but still, any of these must end up in my perspective to be then actualized relative to me.

### 6.4 Einstein-Locality is Fundamental

Einstein-locality implies an upper limit on interaction speeds when they are measured relative to center-of-mass reference frames, namely the velocity of light. Einstein-locality can be an emergent symmetry in solid state systems or on stringy membranes. The standard model is based on relativistic *quantum* field theory. Hence, claiming fundamentality on grounds of the success of the standard model hangs circularly on rejecting violation of Einstein-locality in quantum physics. I have two main reasons for holding Einstein-locality fundamental, apart from those previously presented (Vongehr 2009)[16]:

1) Relative to the fastest interaction itself, for example a photon's reference system, the interaction is instantaneous (on the light cone; having zero eigen-time). The light cone structure therefore provides a "timeless" version where all correlations are instantaneous synchronizations, or in other words, a description that is sufficiently fundamental so that time is apparent. This can be argued necessary in a theory of everything (all times) where time is fully included (called 'parameterized models' – see also below). Light velocity is fundamentally not a certain large value of velocity, but the zero point (no time, so no change, thus no velocity) relative to which the relative velocities between emergent center of mass rest frames become meaningful in the first place.

2) Although one cannot exploit quantum correlations to transport *information* with superluminal velocities, they do require *correlations* to be established with velocities faster than light (if assuming a single world). Such "spooky interaction at a distance" [b] is already incompatible with naïve realism, which includes localism. "Non-local realism" requires precisely the spookiness which realism refutes. This argument was presented in more detail in (Vongehr 2012a)[2].

**6.5  Remarks on Probability**

Lewis (1980)[17] held "objective chance" necessary for probabilistic laws although he held 'subjective Bayesianism' to be more satisfactory. This conflicts with his own view of that 'objective chance' is irrelevant, since such must lead to the epistemic probabilities or is not knowable: Lewis' principal principle demands that objective chances must have the same mathematical structure as Bayesian probabilities. This is symptomatic for that such objective/subjective distinction is verification transcendent. There is model-external and internal randomness, though the internal one involves 'subjective Bayesianism'.

*6.5.1  Model-external Randomness as "Meta-Randomness" Regress Error*

We want *the model* to describe probability. Model-*external* randomness thus must be rejected just like "flow of time": No 'now-moment' creeps along the *t*-axis; there is no "meta-time" outside of fundamental space-time, allowing time inside to flow. Classical general relativity implies cosmology and thus the totality of space of a classical (single world) universe. There is no outside "meta-space" into which cosmic metric expansion penetrates. This is usually accepted with Einstein relativity, at least for the spatial aspect, not as widely for time yet, but again, one usually desires 'fully parameterized' models (~ no external parameters). However, it is the same issue also with Everett relativity in quantum mechanics, which is about the foundation of probability and actualization as the respective parameters; and here the totality of the parameter encompasses the totality of all that is possible, which may be describable as if populated via cosmic quantum inflation producing every possible combination endlessly (just not needing absolute time). Also randomness must be model-*internal*, not outside of totality. Consequently, probability is the biggest problem, called 'measure problem' (Page 2008)[18], in current

---

[b] In 1947, Einstein wrote to Max Born that he could not believe that quantum physics is complete "because it cannot be reconciled with the idea that physics should represent a reality in time and space, free from

cosmology, where also space-in-space and time-of-time misconceptions are most problematic.

Meta-levels, though non-fundamentally allowed (perhaps we are in a 'universe on a membrane'), are not fundamental answers, because they require 'meta-meta' levels to explain the meta-level, leading to regress without definite termination or infinite regress. Consider the branching tree, similar to Figure 7, of the potential outcomes of a coin toss. Model-external randomness assumes two equi*probable* kinds of outcomes (heads or tails), rather than just two kinds. This must be based on indifference due to symmetry, for example both branches having the same number (or other, non circularly defined measure) of sub-worlds, or otherwise the description starts a regress in the following sense. A branch cross section's area seems to be like its phase space volume *C*. Absolute actualization (**ACT**) actualizes the outcome *without bias* for any points in *C*, similar to the ergodic hypothesis of statistical mechanics. The desired randomness is only truly random if **ACT** truly *selects fairly without bias*, which only happens if some further randomness ensures that **ACT** indeed does so! Although it seems natural that a thicker branch will be more likely selected, the randomness is not fully described by the tree alone. This model-external randomness can be called "meta-randomness".

There *could* conceivably be multiple levels of randomness. Overambitious cutting with Occam's (Ockham's) razor leads usually astray. A biased **ACT** perhaps belongs to a set including all possible biases, and I may model none of them, or include one, or all (Figure 2). However, there can fundamentally (by the meaning of "fundamental") not be another meta-level on which we throw another fair 'meta-coin' whenever we reach a branching. Fundamentally, we do not "reach" a branching anyway (no "flow of time"). The branching structure itself describes the randomness. All futures are actualized relative to their own branch. This 'spread of actualization' was here shown via the failure of **ACT** and also (Vongehr 2012a, 2013b)[2,19] by inverting Popper's proof of indeterminism (Popper 1956)[20]. Model-internal observers never select or count branches. Observers remember their past; that is enough. In most worlds, the relative frequency in the empirical record is close to that established by branch counting (3.2), and therefore,

---

spooky actions at a distance."

the modeler counts branches. This is *not* model-external randomness. Our external modeler's perspective changes nothing in the model.

**6.6 What "Realism"?**

*6.6.1 Necessary Fundamental Understanding*

The "real", Latin realis, ~ "of the thing itself", a term invented in the 13th century, is that which "exists absolutely" (absoluteness of being), the "immediate object" of that which is "true". "Real" is also used to signify the "objectively existing" versus the potential, or the physical versus the mental. These are very different meanings, especially the physicalism, because realism originally claims the "reality" of *abstract* and *general* terms (~ ideas, ~ "idealism"), namely of universals, having equal or even superior "reality" to actual physical particulars. "'Objective reality,' which for Descartes signified something mental, has come by a quirk of history to be synonymous with 'extramental existence'." (Kenny1968, p.132)[21] Realism in ontology still is the 'reality of universals' as if they exist "before" things (*ante res*); universals appear *post res* in 'nominalism'. Realism in epistemology can be 'direct/naive realism' about sense experience being true and it is thus closer to (naïve) physicalism. In any case, the "real" is intended to be "absolute", which is itself however relative (6.2). The desired 'absolute absoluteness' can only be had via *self-reference*, which returns to the original "of the thing itself" definition(!): Something may move or exist relative to you and me and is thus "real" in the sense of being inter-subjectively shared for instance, but relative to itself, it is at rest. Relative to me, I cannot but be at rest and exist. This is not trivial, as for example light has itself no reference frame – trying to approach it asymptotically; no time or space remains for it (6.4). Distinguishable Schrödinger cats in quantum superposition exist each relative to itself, via their own experience. The superposition does not have its own perspective and exists relative to the experimenter. A system's existence *relative to itself* is the only "absolute existence" because no regress renders it relative to something else. Totality may be said to exist absolutely because it exists relative to itself via containing (by definition) **everything that anything exists relative to** (this is perspectivism, not a set of all sets!). "Real" as objectively or absolutely existing falls together with the distinction objective/subjective or absolute/relative. Self-reference is the last resort.

### *6.6.2 Realisms and Quantum Mechanics toward Anti-realism*

In Deutsch's interpretation (1997)[22], quantum computing is as powerful as parallel computing, because the computation is distributed over many 'parallel worlds' – it beats classical parallelism via a large number of worlds. Since they all contribute to the result, they are "real" *in that sense*. Quoting Everett (1957)[23] and DeWitt (1973)[24], Kent (1990)[14] remarked that many-worlds interpretations have realism as their main of two essential characteristics. Tegmark (2007)[25] puts the "External Reality Hypothesis (ERH): There exists an external physical reality completely independent of us humans" prominently on the first page, before the mathematical universe hypothesis. These "mathematical realisms" define the "real" implicitly as all that which is necessary in the theory.

'Direct/naïve realism'[26] naively takes how things seem as if directly (without criticism) taken from the senses (there is yet another use of "direct realism" in the philosophical literature). In physics, it supposes that objects with all their properties are a certain definite way "really *out there*", which includes 'localism'(!); "spooky" non-locality already contests naïve realism, as does 'contextuality'. In "local realism", "realism" therefore means 'counterfactual definiteness' (CFD) or 'non-contextuality'. These are both incompatible with quantum experiments, and hence, "*empirically* based modal realism" expresses that alternative situations ("parallel worlds") are labeled "real" because their correlations cannot be omitted from the physical description if it is to predict correctly (like the mentioned "mathematical realisms"). Philosophically prior 'modal relativity' takes modal totality to be by definition the domain of theorizing of a theory of everything possible, because any bias for certain alternatives, such as absolute actualization, is meaningless from the outset, even if this were not yet *empirically* proven. This '*tautological* modal realism' is therefore a form of "anti-realism", namely refusal of verification transcendent distinctions (not necessarily *empirical* verification). The realism/anti-realism distinction is otherwise verification transcendent. This subtlety is characteristic for our transitioning into the modal-relativistic paradigm: Many distinctions are only meaningful in classical models. Like **ACT**, they fail to point anything out in more fundamental descriptions.

Considering quantum gravity, Hawking proposed an "anti-realistic" observer dependent definition of particles already in the seventies, endorsing "something like the Everett-Wheeler interpretation of quantum mechanics" (Gibbons 1977, abstract)[27]. Zeilinger stresses anti-realism for many years, recently with a novel setup (Lapkiewicz 2011)[28]. "Real" always labels what is in some sense inter-subjective. Thus, the label is relative to the identification of subjects. A solipsist's identification may focus merely on Kant's transcendental unity of apperception, the possible states of mind of which include mine and yours in the universal quantum superposition. Black hole complementarity (Susskind 1993)[29] and the 'measure problem' support "local causal patch" descriptions around observers, the volume of which is describable as an illusion suffered by its boundary, a horizon of potentiality (Susskind 1995; Bousso 2006)[30,31]. Relational Quantum Mechanics (Rovelli 1996)[32] completely removes observer independent states from Everett modal realism and resolved EPR[15]. My point of view on modal totality being self-evidently the basis of a ToE that is possible, with unified Einstein-Everett relativity derivable from the modal relativity, is yet another alley of support that independently resolved EPR. In spite of the growing importance of 'anti-realist' modal relativity, even just the successful resolution of the EPR paradox is still widely unknown! We need to show visually intuitive models.